\RequirePackage{lineno}%DELETE FOR PLAIN VERSION
\documentclass[superscriptaddress,aps,preprint,amsmath,amssymb,floatfix]{revtex4}
\usepackage{calrsfs}

\usepackage{graphicx,morefloats}
\usepackage{subfigure} % referring to Figure a and b separately
\usepackage{color,soul}
\usepackage{bm}
\usepackage{amsmath}
\usepackage{amssymb}
\usepackage{times}
\usepackage{hyperref}
\usepackage{bbm}

\newcommand{\YRS}{YbRh$_2$Si$_2$}
\newcommand{\iYRS}{$^{174}$YbRh$_2$Si$_2$}
\newcommand{\LRS}{LuRh$_2$Si$_2$}

\begin{document}
%\linenumbers
%\usepackage{bibunits}%to split refs for main part and Methods

\hyphenation{va-ni-sh-ing li-quid}

\begin{center}

\thispagestyle{empty}

{\large\bf Superconductivity in an extreme strange metal}
\\[0.3cm]

D.\ H.\ Nguyen,$^1$ A.\ Sidorenko,$^1$ M.\ Taupin,$^1$ G.\ Knebel,$^2$ G. Lapertot,$^2$\\  E.~Schuberth,$^3$ and S.~Paschen$^{1,\ast}$\\[0.3cm]

$^1$Institute of Solid State Physics, Vienna University of Technology, 1040
Vienna, Austria\\

$^2$Universit\'e Grenoble Alpes, CEA, Grenoble INP, IRIG, PHELIQS, 38000 Grenoble, France\\

$^3$Technische Universit\"at M\"unchen, 85748 Garching, Germany

\end{center}

%%%%%%%%%%%%%% ABSTRACT %%%%%%%%%%%%%%%%%%%%%%%%%%%%%%%%%%%%%%%%%%%%%%%%%%%%%%%%
\noindent{\bf ABSTRACT}

\noindent{ Some of the highest-transition-temperature superconductors across
various materials classes exhibit linear-in-temperature `strange metal' or
`Planckian' electrical resistivities in their normal state. It is thus believed
by many that this behavior holds the key to unlock the secrets of
high-temperature superconductivity. However, these materials typically display
complex phase diagrams governed by various competing energy scales, making an
unambiguous identification of the physics at play difficult. Here we use
electrical resistivity measurements into the micro-Kelvin regime to discover
superconductivity condensing out of an extreme strange metal state---with linear
resistivity over 3.5 orders of magnitude in temperature. We propose that the
Cooper pairing is mediated by the modes associated with a recently
evidenced dynamical charge localization--delocalization transition, a mechanism
that may well be pertinent also in other strange metal superconductors.}
\vspace{0.5cm}

\noindent E-mail: $^{\ast}$paschen@ifp.tuwien.ac.at
\newpage

%%%%%%%%%%%%%% INTRO %%%%%%%%%%%%%%%%%%%%%%%%%%%%%%%%%%%%%%%%%%%%%%%%%%%%%%%%%%%

\noindent{\bf INTRODUCTION}

\noindent The phenomenon of superconductivity has fascinated scientists since
its discovery in 1911. It had to await microscopic understanding, achieved in
the BCS theory \cite{Bar57.1}, for almost 50 years. Conventional superconductors
such as aluminum and niobium are now key players in the race for realizing the
quantum computer \cite{Wen17.2}. Understanding superconductivity that goes
beyond this framework, as first seen in the the heavy fermion compound
CeCu$_2$Si$_2$ \cite{Ste79.1} and since then considered for numerous materials
classes \cite{Ste17.2}, is the next grand challenge.

In several of these `unconventional' superconductors---across high-$T_{\rm c}$
cuprates \cite{Leg19.1}, iron pnictides and organic conductors \cite{Doi09.1},
heavy fermion metals \cite{Par08.1,Kne08.1} and, very recently, infinite-layer
nickelates \cite{Li19.3}, twisted bilayer graphene \cite{Cao20.1}, and WSe$_2$
\cite{Wan20.1}---the normal state shows a linear-in-temperature `strange metal'
electrical resistivity, at least in certain temperature ranges. This suggests
that to decipher this type of superconductivity requires understanding the
mechanism of the underlying strange metal normal state. However, because of the
complexity of the phase diagrams of many of these systems, this has proven
challenging. Multiple competing effects \cite{Ste17.2}, crossovers between
different scaling behaviors \cite{Mar03.1}, possible trivial
linear-in-temperature resistivity contributions \cite{Pol19.1}, or simply the
fact that strong superconductivity covers much of the normal-state phase space
and needs to be suppressed by external parameters \cite{Pro19.1}, which may
modify the original normal state, are inhibiting consensus on the key mechanism
at play.

On the other hand, there is a material---the heavy fermion compound
YbRh$_2$Si$_2$ \cite{Tro00.2}---where such complications are absent and
linear-in-temperature strange metal behavior
\cite{Tro00.2,Geg02.1,Kne06.1,Tau15.1} has recently been pinned down as
resulting from a dynamical electron localization-delocalization transition
\cite{Pro20.1} as the Kondo effect is destroyed at a magnetic quantum critical
point (QCP) \cite{Si01.1}. Alas, this QCP appeared to lack superconductivity.

Our work unblocks this situation. By developing electrical resistivity
measurements down to record low temperatures---more than 1.5 orders of magnitude
below state-of-the-art---and using them to study YbRh$_2$Si$_2$, we discover
unconventional superconductivity condensing out of a further expanded range of
strange metal behavior, now covering 3.5 orders of magnitude in temperature.
This establishes the connection between electron
localization-delocalization-derived strange metal behavior and
superconductivity, discussed previously for several other materials
\cite{Bal03.1,Bad16.1,Oik15.1,Cao18.1}, to a new level of confidence, thereby
pointing to its universality and putting the spotlight on Cooper pairing
mediated by the critical modes that are associated with this
transition.

We note that in purely itinerant systems at the border of an antiferromagnetic
phase with spin density wave (SDW) order, antiferromagnetic paramagnons
\cite{Sca12.1} may lead to deviations from Fermi liquid behavior (although
generally not to strictly linear-in-temperature resistivities) and provide
superconducting pairing, a mechanism evoked for CePd$_2$Si$_2$  \cite{Mat98.1}.
However, for the above strange metals
\cite{Leg19.1,Doi09.1,Par08.1,Kne08.1,Li19.3,Cao20.1,Wan20.1} this mechanism
seems unlikely, because no magnetic phase exists nearby and/or because there is
evidence that this (weak-coupling) magnetic order parameter description is
inappropriate.

%%%%%%%%%%%%%% THE MATERIAL %%%%%%%%%%%%%%%%%%%%%%%%%%%%%%%%%%%%%%%%%%%%%%%%%%%%

The heavy fermion compound \YRS\ exhibits a low-lying antiferromagnetic phase
that is continuously suppressed to zero by a small magnetic field, establishing
a field-induced QCP \cite{Geg02.1}. A linear temperature dependence of the
electrical resistivity is seen below about 10\,K \cite{Geg02.1,Kne06.1,Tau15.1}.
This behavior is ruled out to be due to electron-phonon scattering because the
non-$f$ reference material \LRS\ is a normal metal and because Fermi liquid
behavior is recovered when \YRS\ is tuned away from the QCP by magnetic field
\cite{Geg02.1,Kne06.1,Tau15.1}. The recent observation of quantum critical
energy-over-temperature ($\omega/T$) scaling in the charge dynamics
\cite{Pro20.1}, together with jumps in the extrapolated zero-temperature Hall
coefficient \cite{Pas04.1,Fri10.2} and associated thermodynamic \cite{Geg07.1}
and spectroscopic signatures \cite{Sei18.1} identifies a dynamical electron
localization--delocalization transition as underlying the strange metal
behavior.

%%%%%%%%%%%%%% RESULTS %%%%%%%%%%%%%%%%%%%%%%%%%%%%%%%%%%%%%%%%%%%%%%%%%%%%%%%%%

%%%%%%%%%%%%%% FIGURE 1 %%%%%%%%%%%%%%%%%%%%%%%%%%%%%%%%%%%%%%%%%%%%%%%%%%%%%%%%

\noindent{\bf RESULTS AND DISCUSSION}

\begin{figure}[t]
\begin{center}
\includegraphics[width=\textwidth]{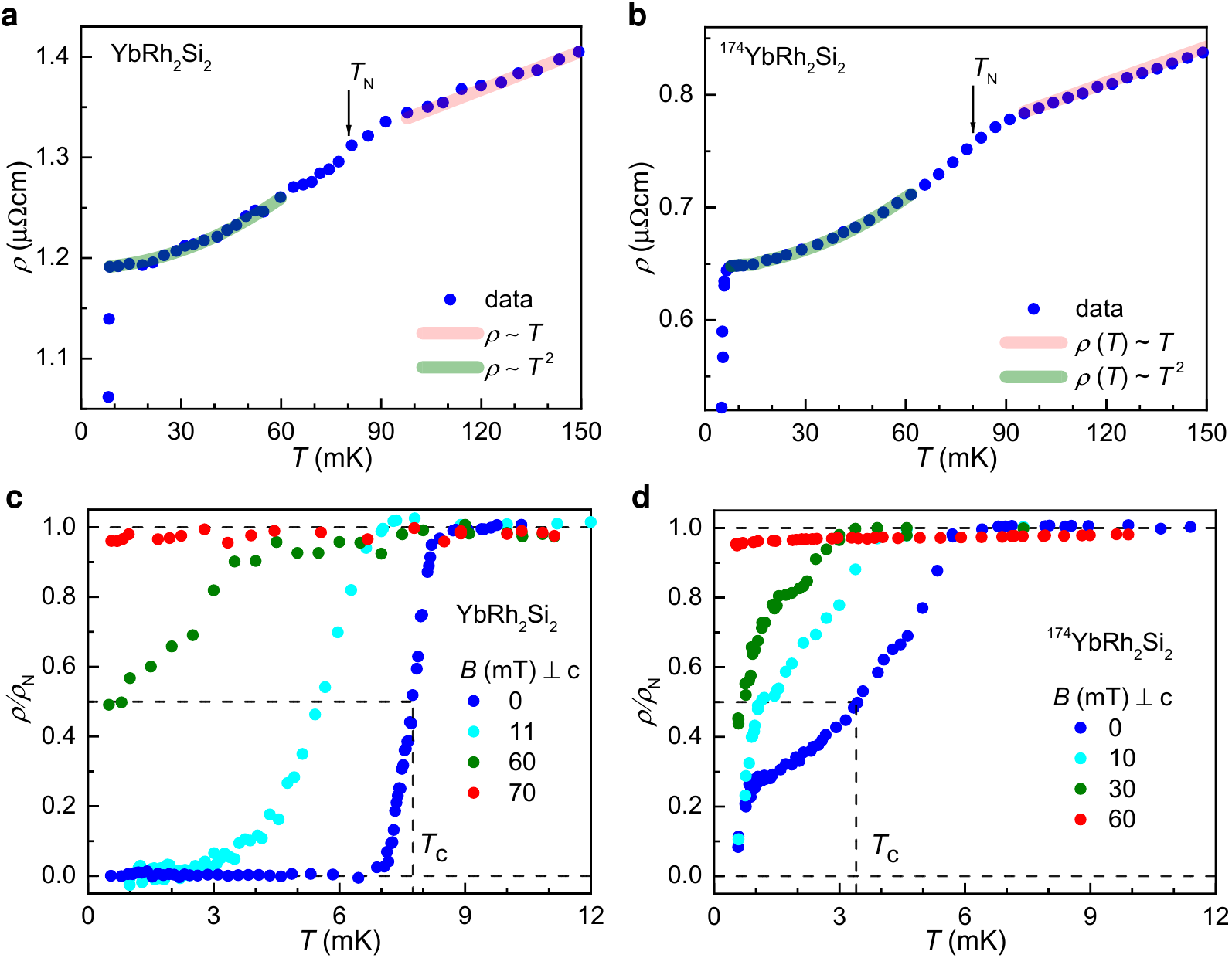}
\end{center}
%\internallinenumbers%DELETE FOR PLAIN VERSION
\caption{
\baselineskip24pt
{\bf Temperature dependence of the electrical resistivity of \YRS\ and
\iYRS.} {\bf a,\,b} Electrical resistivity $\rho(T)$ below 150\,mK, showing
linear-in-$T$ behavior above the N\'eel temperature $T_{\rm N}$, $T^2$ behavior
below it, and the onset of superconductivity at the lowest temperatures. {\bf
c,\,d} Electrical resistivity below 12\,mK, scaled to its normal-state
resistivity $\rho_{\rm N}$ just above the transition, showing the
superconducting transition at $T_{\rm c}$, which we define as the temperature
where $\rho(T)$ has dropped to $\rho_{\rm N}/2$. Magnetic fields (applied within
the tetragonal $a$-$a$ plane) successively suppress $T_{\rm c}$. The 10 and
30\,mT curves for \iYRS\ were extracted from isothermal field sweeps
(Fig.\,\ref{fig3}b); all other curves were recorder as function of temperature.}
\label{fig1}
\end{figure}

We have developed, and implemented in a nuclear demagnetization cryostat
\cite{Ngu12.1}, a setup for high-resolution electrical resistivity measurements
to temperatures well below 1\,mK (see Methods), and used it to measure
state-of-the-art \YRS\ single crystals \cite{Kre12.1,Kne06.1}. To assess the
previously discussed \cite{Sch16.1} role of Yb nuclear moments, in addition to
samples with natural abundance Yb (containing 14.2\% $^{171}$Yb with a nuclear
moment $I=1/2$ and 16.1\% $^{173}$Yb with a nuclear moment $I=5/2$)
\cite{Kre12.1}, we also studied \iYRS, which is free of nuclear Yb moments
\cite{Kne06.1}. In zero magnetic field, both samples show the characteristic
kink near the N\'eel temperature $T_{\rm N}$, as well as linear-in-temperature
behavior above and Fermi liquid $T^2$ behavior below (Fig.\,\ref{fig1}a,b). The
parameters extracted from these fits (Table\,\ref{tab_sample}) are in good
agreement with previous results \cite{Tro00.2,Geg02.1,Kne06.1,Tau15.1},
confirming the high reproducibility of the properties in state-of-the-art \YRS\
single crystals. At the lowest temperatures, a pronounced drop of the
resistivity signals the onset of superconductivity. It is fully displayed in
Fig.\,\ref{fig1}c,d. In \YRS, the transition is rather sharp ($\Delta T_{\rm
c}/T_{\rm c} = 0.10$ for a resistance change from 90\% to 10\% of the value
above the transition), with an onset somewhat below 9\,mK. In \iYRS, the onset
is shifted to below 6\,mK, and the transition is broadened, even though the
residual resistance ratio of this sample is almost twice that of the normal
\YRS\ sample (Table\,\ref{tab_sample}), an effect that will be discussed later.

%%%%%%%%%%%%%% FIGURE 2 %%%%%%%%%%%%%%%%%%%%%%%%%%%%%%%%%%%%%%%%%%%%%%%%%%%%%%%%

\begin{figure}[t]
\begin{center}
\includegraphics[width=\textwidth]{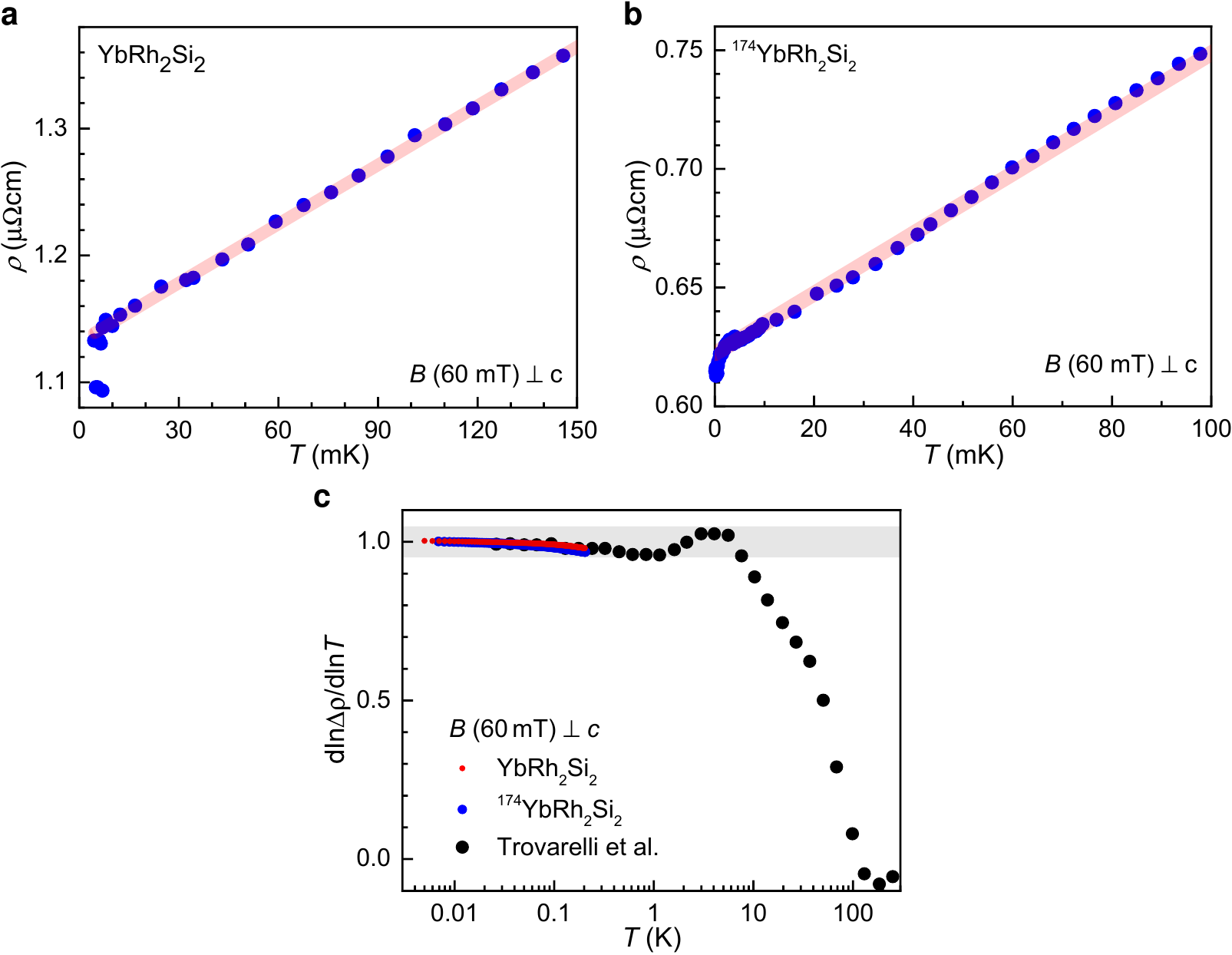}
\end{center}
%\internallinenumbers%DELETE FOR PLAIN VERSION
\caption{
\baselineskip24pt
{\bf Strange metal behavior of \YRS\ and \iYRS.} {\bf a,\,b}
Temperature-dependent electrical resistivity at the quantum critical field of
60\,mT applied within the tetragonal $a$-$a$ plane. Superconductivity develops
out of a strange metal $\rho \sim T$ normal state. The parameters of the
linear-in-$T$ fits are given in Table\,\ref{tab_sample}. {\bf c} Electrical
resistivity exponent $\epsilon$ of the non-Fermi liquid form $\rho = \rho_0' +
A'T^{\epsilon}$, determined as ${\epsilon} ={\rm d}{\ln{\Delta\rho}}/{\rm
d}{\ln{T}}$, shown together with literature data up to high temperatures
\cite{Tro00.2}. This establishes linear-in-$T$ resistivity, evidenced
by $\epsilon = 1 \pm 0.05$, over 3.5 orders of magnitude in temperature. Note that this presentation visualizes the temperature dependence of the exponent with great sensitivity, and that the error bar of $\pm 0.05$ corresponds to a high precision in the exponent's closeness to the value of 1 (for comparison see, e.g., Fig.\,12 of ref.\citenum{Kne08.1}). }
\label{fig2}
\end{figure}

Application of magnetic fields within the $a$-$a$ plane of the tetragonal
crystal successively suppresses superconductivity in both samples
(Fig.\,\ref{fig1}c,d). Note, however, that clear signs of superconductivity are
visible even at the quantum critical field (see Fig.\,\ref{fig1}c, green curve,
for \YRS, and Fig.\,\ref{fig2}b for \iYRS), thus demonstrating superconductivity
that nucleates directly out of the strange metal state (Fig.\,\ref{fig2}a,b).
Our measurements expand the previously established strange metal regime to a
record span of linear electrical resistivity over 3.5 orders of magnitude in
temperature, with a high accuracy of 5\% in the linear exponent $\epsilon$
(Fig.\,\ref{fig2}c).

%%%%%%%%%%%%%% FIGURE 3 %%%%%%%%%%%%%%%%%%%%%%%%%%%%%%%%%%%%%%%%%%%%%%%%%%%%%%%%

\begin{figure}[t!]
\begin{center}
\includegraphics[width=\textwidth]{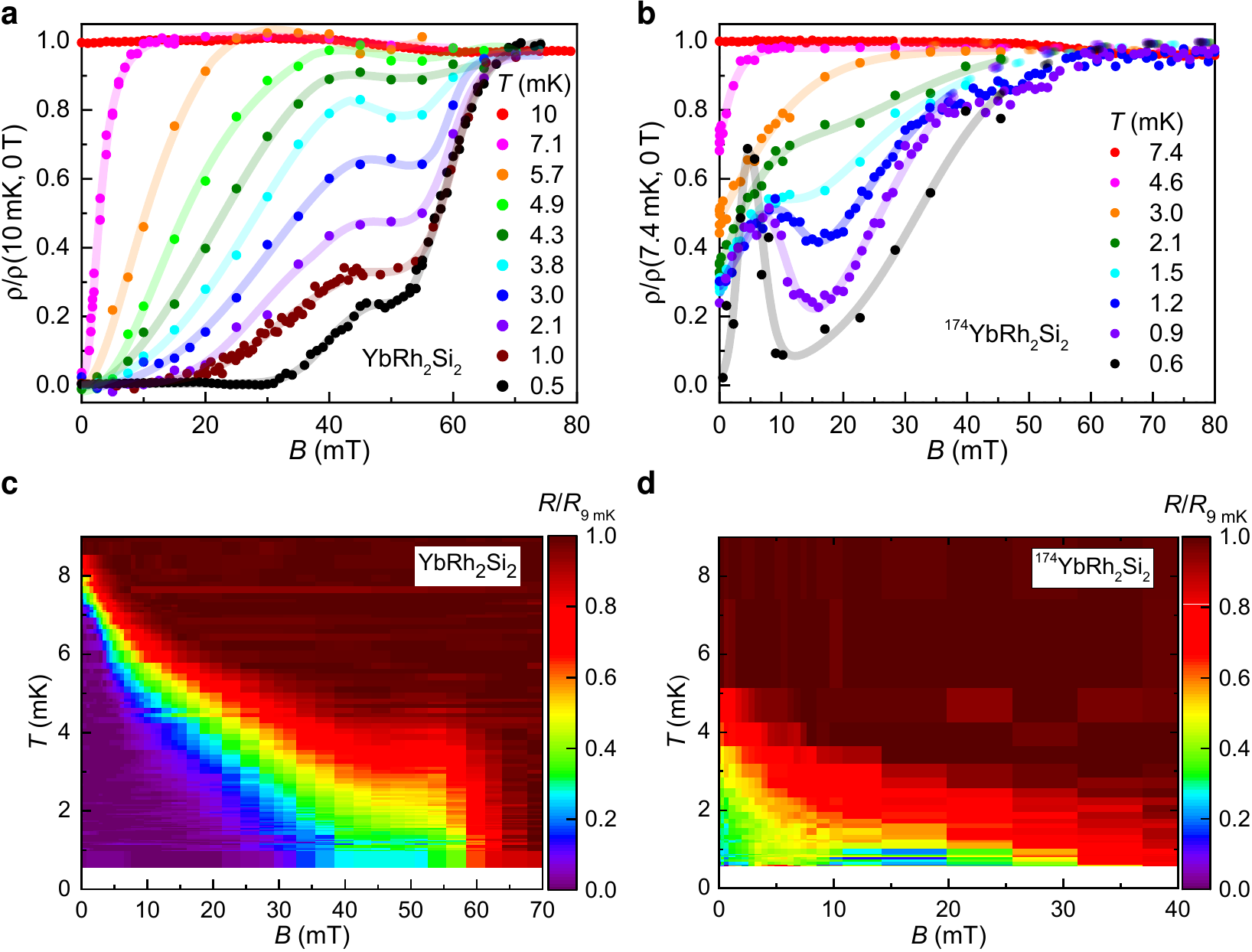}
\end{center}
%\internallinenumbers%DELETE FOR PLAIN VERSION
\caption{
\baselineskip24pt
{\bf Magnetic field dependence of the electrical resistivity of \YRS\ and
\iYRS.} {\bf a,\,b} Selected isotherms of the electrical resistivity $\rho(B)$,
revealing reentrant superconductivity at low temperatures in \iYRS\ ({\bf b})
and remnants thereof in \YRS\ ({\bf a}). Lines are guides-to-the-eyes in {\bf a}
and fits to a multi-transition model (Supplementary Note~1: Analysis of
resistivity vs magnetic field isotherms and Supplementary Fig.\,1) in {\bf b}.
{\bf c,\,d} Temperature--magnetic field phase diagrams of the two samples, with
the reduced electrical resistance $R/R(9\,{\rm mK})$ as color code. The yellow
color corresponds to the resistive midpoint. As a measure of $T_{\rm c}(B)$ it
indicates a phase boundary of unusual shape for \YRS, and two, possibly distinct
phases for \iYRS.}
\label{fig3} 
\end{figure}

To characterize the superconductivity further, we performed isothermal magnetic
field sweeps. In \YRS, all traces of superconductivity disappear only in fields
beyond 70\,mT, which is well above the quantum critical field of 60\,mT
(Fig.\,\ref{fig3}a). The unusual two-step-like shape of the resistivity
isotherms below 5\,mK, which is distinct from normal broadening in applied
fields, suggests that two different superconducting phases might be involved.
This is corroborated by the results on \iYRS, where the lowest-temperature
isotherms show clear signs of reentrance (Fig.\,\ref{fig3}b). We note that no
current dependence was observed, ruling out that flux-flow resistivity is at the
origin of these characteristics. Next, we present color-coded
temperature--magnetic field phase diagrams of \YRS\ and \iYRS\
(Fig.\,\ref{fig3}c,d), constructed from a large number of isotherms. Their merit
is to give a general and fully unbiased impression of the phases present: an
`S-shaped' superconducting region in \YRS\ and two, possibly separated
superconducting regions in \iYRS.

%%%%%%%%%%%%%% FIGURE 4 %%%%%%%%%%%%%%%%%%%%%%%%%%%%%%%%%%%%%%%%%%%%%%%%%%%%%%%%

\begin{figure}[t!]
\begin{center}
\includegraphics[width=\textwidth]{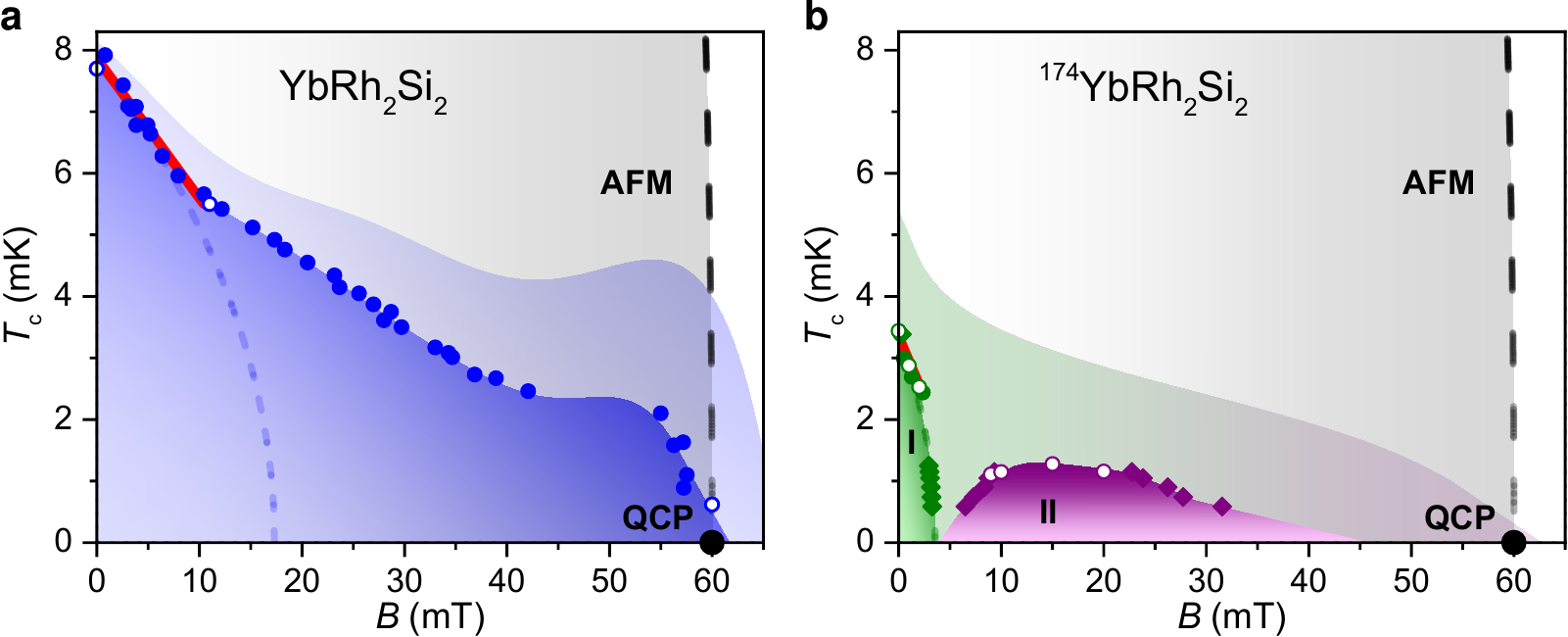}
\end{center}
%\internallinenumbers%DELETE FOR PLAIN VERSION
\caption{
\baselineskip24pt
{\bf Temperature--magnetic field phase diagrams of \YRS\ and \iYRS.}  {\bf a}
Results for \YRS. {\bf b} Results for \iYRS. Open ($T_{\rm c}$) and full
($B_{\rm {c2}}$) circles of all colors represent midpoints (50\%) of the
resistive transitions of iso-$B$ $\rho(T)$ curves (Fig.\,\ref{fig1}c,d) and
iso-$T$ $\rho(B)$ curves (Fig.\,\ref{fig2}a,b), respectively. Full diamonds
represent the $B_{\rm {c2}}$ values extracted from a multi-transition model
(Supplementary Note~1: Analysis of resistivity vs magnetic field isotherms and
Supplementary Fig.\,1) for the three involved transitions (green for leaving
phase I, violet for entering and leaving phase II). The good agreement between
diamonds and circles in the regions of minimal multi-phase interference
justifies this procedure. The upper boundaries of the pale shaded areas away
from data points represent 90\% resistance lines. The black dashed line in both
panels indicates the boundary of the antiferromagnetic (AFM) phase. It
represents an extrapolation of data from refs.\,\citenum{Fri10.2,Geg07.1}, to
$T=0$, the quantum critical point (QCP, black dot). The red lines are linear
fits to the low-field (50\%) data, and determine both the upper critical field
slopes $-d B_{\rm c2}/d T|_{T_{\rm c}}$ and the zero-field superconducting
transition temperatures $T_{\rm c}$ [$= T_{\rm c}(H=0)$]. The dashed lines
represent (inverted) $B_{\rm{c2}}(T)=B_{\rm{c2}}(T=0)\cdot [1-(T/T_{\rm c})^2]$
mean-field curves determined from $-d B_{\rm c2}/d T|_{T_{\rm c}}$ and $T_{\rm
c}$, as estimates of the (low-field) phase boundaries. Below 0.5\,mK, all other
phase/shading boundaries correspond to linear-in-$B$ extrapolations of fits to
the lowest $T$ data points. Note that the perfect overlap of the full and open
symbols demonstrates the high reproducibility of the results.}
\label{fig4} 
\end{figure}

For more quantitative phase diagrams, a definition of the (field-dependent)
transition temperatures $T_{\rm c}$ and (temperature-dependent) upper critical
fields $B_{\rm {c2}}$ has to be adopted. We choose the midpoints of the
resistive transitions, as sketched in Fig.\,\ref{fig1}c,d. For \YRS\ this leads
to the phase boundary delineated by the full ($B_{\rm {c2}}$) and open  ($T_{\rm
c}$) blue circles in Fig.\,\ref{fig4}a, which closely resembles the shape drawn
by the yellow color code (50\% resistance) in Fig.\,\ref{fig3}c. An initial
rapid  suppression of $T_{\rm c}$ is followed by a much more gradual one,
indicating that moving towards the QCP boosts the superconductivity against the
general trend of field suppression associated with the Pauli- and/or
orbital-limiting effect of the magnetic field (for cartoons of this field
effect, see Supplementary Fig.\,3). This is seen even more clearly in the 90\%
resistance line (boundary of the pale shading in Fig.\,\ref{fig4}a), that
exhibits a local maximum at a magnetic field only slightly below the QCP. This
evidences that at least a component of the superconductivity of \YRS\ is
promoted by the same quantum critical fluctuations that are also responsible for
the extreme strange metal behavior---thus anchoring both phenomena to the
material's QCP. That there might indeed be two distinct superconducting phases,
one more readily suppressed by magnetic field and one that is less field
sensitive, receives further support from the phase diagram of \iYRS, presented
next.

Because the resistive transitions have finite widths, they interfere if two or
more phase boundaries are nearby. For \iYRS, where the `unbiased' color-coded
phase diagram already suggests two adjacent phases, we used a simple model to
disentangle their effects (see Supplementary Note~1: Analysis of resistivity vs
magnetic field isotherms and Supplementary Fig.\,1). Indeed, by fitting this
model to the data we find two distinct phases, a low-field one which we denote
as phase I, and a field-induced one which we call phase II (Fig.\,\ref{fig4}b,
see caption for the meaning of the different symbols). Again, we also show the
90\% resistance line as the boundary of the pale shading. Using this criterion,
phase I and II of \iYRS\ grow together into a single superconducting region,
similar to what is observed for \YRS. Conversely, this adds evidence to the
above-proposed two-phase interpretation of the peculiarly shaped superconducting
region of \YRS\ (see cartoons in Supplementary Fig.\,2). Despite the qualitative
similarities between the phase diagrams of \YRS\ and \iYRS, it is clear that
quantitatively, the superconductivity is much weaker in \iYRS. Thus, whereas
nuclear moments---present in \YRS\ but absent in \iYRS---are not a necessary
ingredient to create superconductivity, they do considerably strengthen it.

%%%%%%%%%%%%%% Discussion %%%%%%%%%%%%%%%%%%%%%%%%%%%%%%%%%%%%%%%%%%%%%%%%%%%%%

%%%%%%%%%%%%%% Superconductor characteristics %%%%%%%%%%%%%%%%%%%%%%%%%%%%%%%%%

In what follows we give a few simple estimates of characteristics of the
superconductivity in \YRS\ and \iYRS\ (see Table~\ref{tab_sc}). From the
zero-field $T_{\rm c}$ values (7.9 and 3.4\,mK for \YRS\ and \iYRS,
respectively) and upper critical field slopes $-d B_{\rm c2}/d T|_{T_{\rm c}}$
(4.4 and 2.1\,T/K, much larger than in conventional superconductors), which we
determined from linear fits in Fig.\,\ref{fig4} (see red lines), we estimate the
weak-coupling BCS Ginzburg-Landau coherence lengths $\xi_{\rm GL}$ (97 and
215\,nm). Together with the relevant (non-quantum critical) Sommerfeld
coefficients (see Table\,\ref{tab_sample}) we derive the average Fermi wave
vectors $k_{\rm F}$ (5.2 and 4.8\,nm$^{-1}$) and, with the residual
resistivities $\rho_0$ (see Table~\ref{tab_sample}), the transport scattering
lengths $l_{\rm tr}$ (371 and 976\,nm) and the Ginzburg-Landau penetration
depths $\lambda_{\rm GL}$ (1.8 and 2.0\,$\mu$m). These describe moderately clean
($l_{\rm tr}> \xi_{\rm GL}$), strongly type-II ($\lambda_{\rm GL}/\xi_{\rm
GL}\gg1/\sqrt{2}$) superconductivity. Estimates of the orbital and Pauli
limiting upper critical fields, via $B_{\rm c2}^{\prime} = -0.7 T_{\rm c} d
B_{\rm c2}/d T|_{T_{\rm c}}$ and $B_{\rm p} = 1.764 k_{\rm B}T_{\rm
c}/(\sqrt{2}\mu_{\rm B})$ (24 and 15\,mT for \YRS, and 5 and 6.4\,mT for \iYRS,
respectively) might well be compatible with the phase boundary of phase I in
\iYRS\ and a putative corresponding low-field phase in \YRS. It is clear,
however, that superconductivity in both \YRS\ and \iYRS\ extends to much larger
fields (Fig.\,\ref{fig4}), providing further evidence for the unconventional
nature of the observed superconductivity.

%%%%%%%%%%%%%% Relation Schuberth et al. %%%%%%%%%%%%%%%%%%%%%%%%%%%%%%%%%%%%%%

Next we discuss how our results relate to previous thermodynamic measurements on
\YRS\ with natural abundance Yb, which provided evidence for superconductivity
away from the QCP: Shielding signals were detected, with onsets somewhat below
10\,mK and near 2\,mK, in fields up to 0.055 and 0.418\,mT, respectively
\cite{Sch16.1}, leaving it open whether this superconductivity is related to the
strange metal state of \YRS. The transition temperature we determine from
zero-field electrical resistivity measurements on \YRS\ (with an onset near
9\,mK) is in good agreement with the upper transition (into the B phase)
detected there, identifying it thus with our phase I. The observation of the
lower transition (into the A + sc phase of ref.\citenum{Sch16.1}) could then be
taken as evidence that the two superconducting phases of \YRS\ postulated above 
intersect (see sketch in Supplementary Fig.\,2a). Alternatively, our low-field
superconducting region could comprise both the B and sc phase of
ref.\citenum{Sch16.1}; then our high-field region would be a separate phase not
detected in ref.\citenum{Sch16.1} (see sketch in Supplementary Fig.\,2b), just
as the high-field phase of \iYRS. This should be clarified by future
magnetization/susceptibility measurements in lower fields (below the background
field of 0.012\,mT reached in ref.\citenum{Sch16.1}, which appears to be well
above the lower critical field of the B phase), ideally on powdered samples to
better assess the Meissner volume of the B phase.

The lower transition of ref.\citenum{Sch16.1} is accompanied by a specific heat
anomaly that releases a giant amount of entropy. It was interpreted as phase
transition into a state of hybrid nuclear and electronic spin order of Yb, which
reduces the internal (staggered) magnetic field of the primary electronic order
and thus creates a less hostile environment for superconductivity. As our study
reveals, the superconductivity in \iYRS, which lacks nuclear Yb moments and can
thus not exhibit such hybrid order, is indeed considerably weaker than in \YRS.
This confirms that nuclear Yb moments boost the superconductivity in \YRS. That
this mechanism apparently works up to temperatures well above the hybrid
ordering temperature may be ascribed to short-range fluctuations, evidenced by
the entropy release extending up to about 10\,mK. Apart from the overall
weakening of the superconductivity in \iYRS\ compared to \YRS, the phase
diagrams of the two materials do, however, share many similarities
(Fig.\,\ref{fig4}), calling for an understanding within the same framework.

%%%%%%%%%%%%%% Theoretical description? %%%%%%%%%%%%%%%%%%%%%%%%%%%%%%%%%%%%%%%

%\noindent{\bf DISCUSSION}

\noindent The question that then arises is what is the mechanism of the Cooper pairing in the detected superconducting phases?

We start by recapitulating our results that make the BCS mechanism extremely
unlikely:  (i) Superconductivity in \YRS\ condenses out of an extreme strange
metal state, with linear-in-temperature resistivity right down to the onset of
superconductivity (Fig.\,\ref{fig2}a); (ii) the upper critical field slope
(Table\,\ref{tab_sc}) as well as the directly measured critical field
(Fig.\,\ref{fig4}a) strongly overshoot both the Pauli and the orbital limiting
fields; (iii) the low-temperature resistivity isotherms exhibit a two-step
transition (Fig.\,\ref{fig3}a), evidencing that one component is much less
field-sensitive than the other; (iv) the superconducting phase boundary deviates
strongly from a mean-field shape (Figs.\,\ref{fig3}a and \ref{fig4}a),
evidencing that the field boosts (at least part of) the superconductivity
against the general trend of field suppression; (v) superconductivity is
strongly suppressed by substituting the natural abundance Yb (of atomic mass
173.04) by $^{174}$Yb, though the isotope effect in a BCS picture would have a
minimal effect (a reduction of $T_{\rm c}$ by 0.1\%). It is thus natural to
assume that quantum critical fluctuations are involved in stabilizing (at least
the high-field part of) the superconductivity in \YRS.

Very recently, in a two-impurity Anderson model that features Kondo destruction
\cite{Si01.1}, the singlet-pairing susceptibility was found to be strongly
enhanced by critical local moment fluctuations \cite{Cai20.1}. Because singlet
pairing may be subject to Pauli limiting, this phase---though stabilized by
quantum critical fluctuations---might be suppressed by the applied magnetic
field even well before the QCP is reached. Thus, phase I of \iYRS\ and the
putative corresponding low-field phase of \YRS\ are promising candidates for
this type of superconductivity. Interestingly, in this context, unconventional
superconductivity is also discussed in the spin-triplet channel
\cite{Kan18.1,Li19.2}. It is tempting to consider phase II of \iYRS\ and the
putative corresponding high-field phase of \YRS\ to be of this kind, which would
provide a compelling link to recently discovered candidate spin-triplet
topological superconductors \cite{Zha18.2,Ran19.1}. Of course, any model for
superconductivity in \YRS\ (and \iYRS) should also correctly capture the normal
state properties.

At the QCP of \YRS, a dynamical electron localization--delocalization
transition, featuring both linear-in-temperature dc resistivity, and
linear-in-frequency and linear-in-temperature `optical resistivity', was
evidenced by quantum critical frequency-over-temperature scaling, with a
critical exponent of 1, of the THz conductivity of \YRS\ \cite{Pro20.1}. This
evidences critical modes in addition to the ones associated
with the suppression of the (antiferro)magnetic order parameter. A microscopic
mechanism compatible with this scaling is the disentanglement of the
(electronic) Yb $4f$ moments from the conduction electrons as static Kondo
screening breaks down at the QCP
\cite{Si01.1,Pro20.1,Pas04.1,Fri10.2,Geg07.1,Sei18.1}. Whether spin-triplet
superconductivity may arise in models that capture this physics remains to be
clarified by future work. Given that the quantum critical magnetic field is
considerably larger than the scale associated with the superconducting
transition temperature near the QCP, a promising direction is to consider the
role of the applied magnetic field as reducing the spin symmetry from being
in-plane continuous to Ising-anisotropic. In fact, calculations in an
Ising-anisotropic two-impurity Bose-Fermi Anderson model suggest that near the
model's Kondo-destruction QCP the spin-triplet pairing correlation is
competitive with the spin-singlet one \cite{Pix15.3}. Naturally, the proposal of
spin-triplet superconductivity should also be scrutinized by future experiments,
including probes of anisotropies and NMR investigations, which are in principle
feasible at ultralow temperatures. 

%%%%%%%%%%%%%% Conclusion %%%%%%%%%%%%%%%%%%%%%%%%%%%%%%%%%%%%%%%%%%%%%%%%%%%%%

Our observation of unconventional superconductivity condensing out of the
textbook strange metal state of \YRS\ ends a debate about reasons for its
(previously supposed) absence. Thus, neither the $4f$ moment stemming from Yb or
the tuning parameter being magnetic field (as opposed to the typical situation
of Ce-based systems under pressure or doping tuning), nor the QCP being governed
by effects  beyond order parameter fluctuations inhibit superconductivity.
Instead, we propose that critical modes associated with a dynamical
electron localization--delocalization transition mediate the strange-metal
unconventional superconductivity in \YRS. Future experiments, ideally in
conjunction with ab initio-based theoretical work, shall ascertain this
assignment, disentangle the different superconducting phases, determine the
symmetry of the order parameter, clarify further important details such as the
single vs multiband nature of the superconductivity, and even explore the
possibility of exotic surface phases.

%%%%%%%%%%%%%% Implications %%%%%%%%%%%%%%%%%%%%%%%%%%%%%%%%%%%%%%%%%%%%%%%%%%%

Finally, we relate our discovery to strange-metal unconventional superconductors
in other materials classes. It has been pointed out that the strange metal
behavior in many of these is compatible with `Planckian dissipation', i.e., with
the transport scattering rate $1/\tau$ in a simple Drude conductor being equal
to $\alpha k_{\rm B}T/\hbar$, with $\alpha \approx 1$ \cite{Bru13.1}. For \YRS\
and \iYRS, we estimate much smaller $\alpha$ values (see Table\,\ref{tab_sc},
and Supplementary Note~2: Estimates on Planckian dissipation), suggesting that
linear-in-temperature resistivity---even in the extreme form observed
here---does not require the Planckian limit to be reached. Whether this is
related to the very strongly correlated nature of this compound (with effective
masses above $1000\cdot m_0$, much beyond those in the materials considered in
ref.\,\citenum{Bru13.1}) or its pronounced deviation from Drude-like behavior
(see Supplementary Materials of ref.\,\citenum{Pro20.1}), is an interesting
question for future studies. As to a microscopic understanding of the
phenomenon, we point out that charge delocalization transitions, similar to those
observed in \YRS, are being discussed also in other strange-metal
superconductors, including the high-$T_{\rm c}$ cuprates \cite{Bal03.1,Bad16.1},
an organic conductor \cite{Oik15.1} and, tentatively, even magic angle bilayer
graphene \cite{Cao18.1}. Our results thus point to the exciting possibility that
a dynamical electron localization--delocalization transition may mediate
strange-metal unconventional superconductivity in a broad range of materials
classes. \vspace{0.5cm}

%%%%%%%%%%%%%%%%%%%%%%%%%%%%%%%%%%%%%%%%%%%%%%%%%%%%%%%%%%%%%%%%%%%%%%%%%%%%%%%%

%%%%%%%%%%%%%%%%%%%%%%%%%%%%%%%%%%%%%%%%%%%%%%%%%%%%%%%%%%%%%%%%%%%%%%%%%%%%%%%
%%%%%%% FIGURE 1 %%%%%%%%%%%%%%%%%%%%%%%%%%%%%%%%%%%%%%%%%%%%%%%%%%%%%%%%%%%%%%
%%%%%%%%%%%%%%%%%%%%%%%%%%%%%%%%%%%%%%%%%%%%%%%%%%%%%%%%%%%%%%%%%%%%%%%%%%%%%%%

%%%%%%%%%%%%%%%%%%%%%%%%%%%%%%%%%%%%%%%%%%%%%%%%%%%%%%%%%%%%%%%%%%%%%%%%%%%%%%%
%%%%%%% FIGURE 2 %%%%%%%%%%%%%%%%%%%%%%%%%%%%%%%%%%%%%%%%%%%%%%%%%%%%%%%%%%%%%%
%%%%%%%%%%%%%%%%%%%%%%%%%%%%%%%%%%%%%%%%%%%%%%%%%%%%%%%%%%%%%%%%%%%%%%%%%%%%%%%

%%%%%%%%%%%%%%%%%%%%%%%%%%%%%%%%%%%%%%%%%%%%%%%%%%%%%%%%%%%%%%%%%%%%%%%%%%%%%%%
%%%%%%% FIGURE 3 %%%%%%%%%%%%%%%%%%%%%%%%%%%%%%%%%%%%%%%%%%%%%%%%%%%%%%%%%%%%%%
%%%%%%%%%%%%%%%%%%%%%%%%%%%%%%%%%%%%%%%%%%%%%%%%%%%%%%%%%%%%%%%%%%%%%%%%%%%%%%%

%%%%%%%%%%%%%%%%%%%%%%%%%%%%%%%%%%%%%%%%%%%%%%%%%%%%%%%%%%%%%%%%%%%%%%%%%%%%%%%
%%%%%%% FIGURE 4 %%%%%%%%%%%%%%%%%%%%%%%%%%%%%%%%%%%%%%%%%%%%%%%%%%%%%%%%%%%%%%
%%%%%%%%%%%%%%%%%%%%%%%%%%%%%%%%%%%%%%%%%%%%%%%%%%%%%%%%%%%%%%%%%%%%%%%%%%%%%%%

%%%%%%%%%%%%%%%%%%%%%%%%%%%%%%%%%%%%%%%%%%%%%%%%%%%%%%%%%%%%%%%%%%%%%%%%%%
%%%%%%%% Methods %%%%%%%%%%%%%%%%%%%%%%%%%%%%%%%%%%%%%%%%%%%%%%%%%%%%%%%%%
%%%%%%%%%%%%%%%%%%%%%%%%%%%%%%%%%%%%%%%%%%%%%%%%%%%%%%%%%%%%%%%%%%%%%%%%%%
%\pagenumbering{arabic}
%\setcounter{page}{1}

%\refstepcounter{SM}{35}
%\setcounter{SM}{35}
%\addtocounter{SM}{35} 

%\newpage

{\noindent\bf METHODS}

\noindent{\bf A. Refrigerator and thermometry} 

\noindent All measurements were carried out in the Vienna nuclear
demagnetization refrigerator \cite{Ngu12.1}. We use resistance thermometry (Pt
resistance thermometers Pt-1000, RuO$_2$ thermometers, and sliced Speer carbon
resistance thermometers) for temperatures above 10\,mK and magnetic thermometry
(CMN, pulsed Pt-NMR thermometers, via the pulsed nuclear magnetic resonance of
$^{195}$Pt nuclei) for temperature below 20\,mK. For the presented results, we
used a CMN thermometer between 20 and 2\,mK, a Speer thermometer above, and a
Pt-NMR thermometer below.

\noindent{\bf B. Filters} 

\noindent To attenuate radiofrequency radiation, a series of filters and
thermalization stages consisting of thermo-coax cables (from room temperature to
the 50\,mK plate), silver-epoxy filters, and RC filters were installed. The
silver-epoxy filters are also used for thermalization (down to the mixing
chamber temperature).

\noindent{\bf C. Sample holders and thermalization} 

\noindent Machined silver sample holders made from a 5N silver rod were annealed
to reach a residual resistance ratio of 2000. They are directly screwed to the
nuclear stage using home-made silver screws. To thermalize the samples, one of
the two outer contact points (of the 4-point technique, see point E below) was
connected by $50\,\mu$m gold wires to the silver holder, via spot welding on the
sample and screwing to the silver holder. A separate grounding point of
excellent quality (anchored 600\,m below ground level) was used. The different
stages of the cryostat (4\,K and 1\,K plate, still, 50\,mK plate, mixing
chamber, copper nuclear stage) were all connected via NbTi/CuNi superconducting
wires to the same ground, such that the ground potential was highly uniform. To
avoid ground loops, all measurements devices were connected via opto-couplers.
During the measurements at ultralow temperatures (below 10\,mK), all measurement
devices were powered by $4 \times 12$\,V$-150$\,A batteries with floating
ground. For electrical isolation (away from the ground point), Vespel was used
on the sample holder, the superconducting coil, and the Nb-superconducting
shield. 

\noindent{\bf D. Magnetic field applied to the samples} 

\noindent A dc magnetic field coil made of superconducting NbTi wire ($T_{\rm c}
= 9.2$\,K), a superconducting Nb cylinder (20\,mm diameter, 10\,cm length), and
the sample holder are concentrically assembled. The field coil and the
superconducting shield are thermalized to the mixing chamber, the sample holder
to the nuclear stage. Magnetic fields at the position of the samples up to
80\,mT were generated by applying a dc current, with the precision of a few
$\mu$T. For the highest fields, a current above 1.4\,A was passed through the
self-made coil. Because this represented a considerable risk of quenching and
breaking the magnet, only a final set of experiments was done up to the highest
fields. In particular, most of the measurements on \iYRS\ were done up to 45\,mT
only.

\noindent{\bf E. Resistivity measurements} 

\noindent Electrical resistivity measurements were done with a 7124 Precision
Lock-in Amplifier. Electrical currents were sourced by a CS580
voltage-controlled current source. The lowest measurement current in our
experiments was 10\,nA. To increase the signal to noise ratio, low-temperature
transformers (encapsulated in a lead shield) with a gain of 1000 were used. They
were installed at and thermalized to the mixing chamber. In addition, a SR560
low noise voltage pre-amplifier was used at room temperature. Electrical
contacts for these measurements where made by spot-welding gold wires to the
samples, in a standard 4-point geometry. Carefully derived measurement protocols
(electrical current densities, sweep rates, waiting times etc.) for a good
thermalization of the samples to the nuclear stage were followed for all
presented measurements; the good thermalization is confirmed by the
reproducibility of the results between different types of experiments (e.g.\
temperature vs magnetic-field sweeps, cooling vs warming curves). Applying
larger currents leads to overheating effects at the lowest temperatures, but no
evidence for flux-flow resistance could be revealed.  \vspace{0.5cm}

%%%%%%%%%%%%%%%%%%%%%%%%%%%%%%%%%%%%%%%%%%%%%%%%%%%%%%%%%%%%%%%%%%%%%%%%%%%%%%%
%%%%%%%%%% OTHERS %%%%%%%%%%%%%%%%%%%%%%%%%%%%%%%%%%%%%%%%%%%%%%%%%%%%%%%%%%%%%
%%%%%%%%%%%%%%%%%%%%%%%%%%%%%%%%%%%%%%%%%%%%%%%%%%%%%%%%%%%%%%%%%%%%%%%%%%%%%%%

{\noindent\large\bf Data availability}\\
The datasets generated during and/or analysed during the current study are available from the corresponding author on reasonable request.
\vspace{0.5cm}

%%%%%%%%%%%%%%%%%%%%%%%%%%%%%%%%%%%%%%%%%%%%%%%%%%%%%%%%%%%%%%%%%%%%%%%%%%%%%%%
%%%%%%%%%% REFERENCES %%%%%%%%%%%%%%%%%%%%%%%%%%%%%%%%%%%%%%%%%%%%%%%%%%%%%%%%%
%%%%%%%%%%%%%%%%%%%%%%%%%%%%%%%%%%%%%%%%%%%%%%%%%%%%%%%%%%%%%%%%%%%%%%%%%%%%%%%

%\bibliographystyle{naturemagallauthors}
%\bibliography{../silke_all}
%\vspace{0.5cm}

\vspace{0.5cm}

%%%%%%%%%%%%%%%%%%%%%%%%%%%%%%%%%%%%%%%%%%%%%%%%%%%%%%%%%%%%%%%%%%%%%%%%%%%%%%%
%%%%%%%%%% MORE OTHERS %%%%%%%%%%%%%%%%%%%%%%%%%%%%%%%%%%%%%%%%%%%%%%%%%%%%%%%%
%%%%%%%%%%%%%%%%%%%%%%%%%%%%%%%%%%%%%%%%%%%%%%%%%%%%%%%%%%%%%%%%%%%%%%%%%%%%%%%

{\noindent\large\bf Acknowledgements}\\
The authors thank A.\ Casey, Ch.\ Enss, G.\ Frossatti, L.\ Levitin, J.\
Saunders, A.\ de Waard, D.\ Zumbuehl and other colleagues from the European 
Microkelvin Platform (EMP) for sharing expertise in ultralow-temperature
techniques, C.\ Krellner for growing \YRS\ single crystals under supervision of
Ch.\ Geibel and F.\ Steglich in Dresden, M.\ Brando for assistance in the
selection of suitable single crystals, and P.\ Buehler, A.\ Prokofiev, Q.\ Si,
and F.\ Steglich for fruitful discussions. The team in Vienna acknowledges
financial support from the Austrian Science Fund (FWF grants P29296-N27 and DK
W1243), the EMP (H2020 Project 824109), and the European Research Council (ERC
Advanced Grant 227378).
\vspace{0.5cm}

{\noindent\large\bf Author contributions}\\
S.P.\ initiated and lead the study. G.K.\ and G.L.\ synthesized and
characterized the \iYRS\ single crystals. D.H.N.\ and A.S.\ set up the
ultralow-temperature experiments, with initial guidance by E.S., and performed
the measurements. D.H.N., M.T., and S.P.\ analyzed the data. S.P.\ wrote the
manuscript, with input from all authors. All authors contributed to the
discussion.
\vspace{0.5cm}

{\noindent\large\bf Competing interests}\\
The authors declare no competing interests.
\vspace{0.5cm}

\noindent{\bf Correspondence} and requests for materials should be addressed to S.P.

%%%%%%%%%%%%%%%%%%%%%%%%%%%%%%%%%%%%%%%%%%%%%%%%%%%%%%%%%%%%%%%%%%%%%%%%%%%%%%%

\newpage

%%%%%%%%%%%%%%%%%%%%%%%%%%%%%%%%%%%%%%%%%%%%%%%%%%%%%%%%%%%%%%%%%%%%%%%%%%%%%%%
%%%%%%% TABLE 1 %%%%%%%%%%%%%%%%%%%%%%%%%%%%%%%%%%%%%%%%%%%%%%%%%%%%%%%%%%%%%%%
%%%%%%%%%%%%%%%%%%%%%%%%%%%%%%%%%%%%%%%%%%%%%%%%%%%%%%%%%%%%%%%%%%%%%%%%%%%%%%%

\noindent{\bf\large Figures and tables}

\begin{table}[!h]
%\internallinenumbers%DELETE FOR PLAIN VERSION
\caption{
\baselineskip24pt
{\bf Characteristics of the investigated \YRS\ and \iYRS\ single crystals.} Both samples are from batches studied in detail previously \cite{Kre12.1,Kne06.1}. Their residual resistance ratios RRR$_{\rm 10\,mK} = R(300$\,K)$/R(10$\,mK), as well as the zero-field Fermi liquid behavior $\rho = \rho_0 + AT^2$ below $T_{\rm N}$ and the non-Fermi liquid behavior $\rho = \rho_0' + A'T$ at the quantum critical field of 60\,mT confirm high sample quality. To remove uncertainties in the geometric factors, we have assumed $\rho(300$\,K$) = 80\,\mu\Omega$cm \cite{Tro00.2}. The Sommerfeld coefficient in zero field $\gamma_{\rm KW}^{\rm 0\,T}$, calculated from $A$ via the universal Kadowaki-Woods ratio $A/\gamma^2 = 10^{-5}\,\mu\Omega{\rm cm}\rm{(mol K)}^2/\rm{(mJ)}^2$, is a good estimate of the non-quantum critical contribution (see Supplementary Note~2: Estimates on Planckian dissipation).}
\vspace{0.5cm}
\small

\hspace{-1cm}\begin{tabular}{| c || c | c | c | c | c | c | c |} 
\hline
sample & batch & RRR$_{\rm 10\,mK}$ & $\rho_0$ ($\mu\Omega$cm) & $A$ ($\mu\Omega$cm/K$^2$) & $\rho_0'$ ($\mu\Omega$cm) & $A'$ ($\mu\Omega$cm/K) & $\gamma_{\rm KW}^{\rm 0\,T}$ (J/molK$^2$)\\ [0.5ex] 
 \hline\hline
 \YRS\ & 63113\_1 & 67 & 1.19 & 20.2 & 1.23 & 1.17 & 1.42\\ 
 \hline
 \iYRS\ & Lap0288 & 123 & 0.55 & 14.8 & 0.59 & 0.85 & 1.22\\
 \hline
\end{tabular}
\label{tab_sample}
\end{table}

%%%%%%%%%%%%%%%%%%%%%%%%%%%%%%%%%%%%%%%%%%%%%%%%%%%%%%%%%%%%%%%%%%%%%%%%%%%%%%%
%%%%%%% TABLE 2 %%%%%%%%%%%%%%%%%%%%%%%%%%%%%%%%%%%%%%%%%%%%%%%%%%%%%%%%%%%%%%%
%%%%%%%%%%%%%%%%%%%%%%%%%%%%%%%%%%%%%%%%%%%%%%%%%%%%%%%%%%%%%%%%%%%%%%%%%%%%%%%

\begin{table}[!h]
%\internallinenumbers%DELETE FOR PLAIN VERSION
\caption{
\baselineskip24pt
{\bf Characteristics of the superconducting phase(s) in \YRS\ and \iYRS.} Both $T_{\rm c}(B)$ and $B_{\rm c2}(T)$ are defined as the midpoints of the resistive transitions (see Figs.\,\ref{fig1}c,d and \ref{fig2}a,b). The zero-field values $T_{\rm c}$ [$= T_{\rm c}(H=0)$] as well as the upper critical field slopes $-d B_{\rm c2}/d T|_{T_{\rm c}}$ are determined from linear fits to the data at small fields (red lines in Fig.\,\ref{fig4}). Listed are also estimates of the Ginzburg-Landau coherence length $\xi_{\rm GL}$, the average Fermi wavevector $k_{\rm F}$, the transport scattering length $l_{\rm tr}$, the Ginzburg-Landau penetration depth $\lambda_{\rm GL}$, the (orbital limiting) upper critical field $B_{\rm c2}^{\prime}$, and the Pauli limiting field $B_{\rm p}$ (see text). $\alpha$ is the prefactor of the scattering rate $1/\tau = \alpha k_{\rm B}T/\hbar$ of a simple Drude conductor, estimated from $T_{\rm c}$, $\xi_{\rm GL}$ (both below), and $\gamma_{\rm KW}^{\rm 0\,T}$ (from Table\,\ref{tab_sample}), as explained in the Supplementary Note~2: Estimates on Planckian dissipation; `Planckian dissipation' refers to $\alpha = 1$.}
\vspace{0.5cm}
\small

\hspace{-1.5cm}\begin{tabular}{| c | c | c | c | c | c | c || c | c || c |} 
 \hline
 sample & $T_{\rm c}$ (mK) & $\frac{-d B_{\rm c2}}{d T}|_{T_{\rm c}}$ ($\frac{\rm T}{\rm K}$) & $\xi_{\rm GL}$ (nm) & $k_{\rm F}$ ($\frac{1}{\rm nm}$) & $l_{\rm tr}$ (nm) & $\lambda_{\rm GL}$ ($\mu$m) & $B_{\rm c2}^{\prime}$ (mT) & $B_{\rm p}$ (mT) & $\alpha$\\ [0.5ex] 
 \hline\hline
 \YRS\ & 7.9 & 4.4 & 97 & 5.2 & 371 & 1.8 & 24 & 15 & 0.011\\ 
 \hline
 \iYRS\ & 3.4 & 2.1 & 215 & 4.8 & 976 & 2.0 & 5.0 & 6.4 & 0.0065\\
 \hline
\end{tabular}
\label{tab_sc}
\end{table}

\newpage

%%%%%%%%%%%%%%%%%%%%%%%%%%%%%%%%%%%%%%%%%%%%%%%%%%%%%%%%%%%%%%%%%%%%%%%%%%%%
%%%%%%%%%%% FIGURES %%%%%%%%%%%%%%%%%%%%%%%%%%%%%%%%%%%%%%%%%%%%%%%%%%%%%%%%
%%%%%%%%%%%%%%%%%%%%%%%%%%%%%%%%%%%%%%%%%%%%%%%%%%%%%%%%%%%%%%%%%%%%%%%%%%%%

%%%%%%%%%%% FIGURE 1 %%%%%%%%%%%%%%%%%%%%%%%%%%%%%%%%%%%%%%%%%%%%%%%%%%%%%%%%

\newpage

%%%%%%%%%%% FIGURE 2 %%%%%%%%%%%%%%%%%%%%%%%%%%%%%%%%%%%%%%%%%%%%%%%%%%%%%%%%

\clearpage

\newpage

%%%%%%%%%%% FIGURE 3 %%%%%%%%%%%%%%%%%%%%%%%%%%%%%%%%%%%%%%%%%%%%%%%%%%%%%%%%

\clearpage

\newpage

%%%%%%%%%%% FIGURE 4 %%%%%%%%%%%%%%%%%%%%%%%%%%%%%%%%%%%%%%%%%%%%%%%%%%%%%%%%

\clearpage

\newpage

%%%%%%%%%%%%%%%%%%%%%%%%%%%%%%%%%%%%%%%%%%%%%%%%%%%%%%%%%%%%%%%%%%%%%%%%%%
%%%%%%%% Supplementary Information %%%%%%%%%%%%%%%%%%%%%%%%%%%%%%%%%%%%%%%
%%%%%%%%%%%%%%%%%%%%%%%%%%%%%%%%%%%%%%%%%%%%%%%%%%%%%%%%%%%%%%%%%%%%%%%%%%
%\pagenumbering{arabic}
\setcounter{page}{1}

%\refstepcounter{SM}{35}
%\setcounter{SM}{35}
%\addtocounter{SM}{35} 

\addtocounter{figure}{-4} 
%\addtocounter{equation}{-3} 
\begin{center}

\noindent{\bf\LARGE Supplementary Information}
\vspace{0.5cm}

{\large\bf Superconductivity in an extreme strange metal}
\\[0.3cm]

D.\ H.\ Nguyen,$^1$ A.\ Sidorenko,$^1$ M.\ Taupin,$^1$ G.\ Knebel,$^2$ G. Lapertot,$^2$\\  E.~Schuberth,$^3$ and S.~Paschen$^{1,\ast}$\\[0.3cm]

$^1$Institute of Solid State Physics, Vienna University of Technology, 1040
Vienna, Austria\\

$^2$Universit\'e Grenoble Alpes, CEA, IRIG, PHELIQS, 38000 Grenoble, France\\

$^3$Technische Universit\"at M\"unchen, 85748 Garching, Germany
\vspace{0.5cm}

\end{center}

\noindent{\bf Supplementary Note~1: Analysis of resistivity vs magnetic field isotherms}

\noindent In \iYRS\ the electrical resistivity vs magnetic field isotherms
(Fig.\,\ref{fig3}b) at the lowest temperatures show clear signs of reentrance.
To extract meaningful upper critical field $B_{\rm{c2}}$ and critical
temperature $T_{\rm c}$ values is thus more involved than in the case of a
single superconductor--normal conductor phase boundary (where the criterion of a
50\% resistivity drop is frequently adopted). We have thus devised a simple
`multi-transition model' to separate the different effects. A single magnetic
field-driven transition from a superconductor to a normal conductor is
(phenomenologically) described by
\vspace{-0.5cm}

\begin{equation}
\rho/\rho_{\rm n} = \frac{1}{1+e^{-k(B-B_{\rm{c2}})}} \equiv \mathcal{F}(B)
\end{equation}

\noindent where $B_{\rm{c2}}$ represents the half-hight (50\% resistivity drop) criterion and $k$ is a measure of the sharpness of the transition. The effect of first entering and then leaving a superconducting phase II as function of magnetic field is then captured by
\vspace{-0.5cm}

\begin{equation}
\rho/\rho_{\rm n} = 1-(1-\mathcal{F}_{\rm left}) \cdot (1-\mathcal{F}_{\rm right}) = \mathcal{F}_{\rm II}(B) \; ,
\end{equation}

\noindent and the total effect of starting in a superconducting phase I, leaving it, and passing through the superconducting phase II by
\vspace{-0.5cm}

\begin{equation}
\rho/\rho_{\rm n} = \mathcal{F}_{\rm I}(B) \cdot \mathcal{F}_{\rm II}(B) \; ,
\end{equation}

\noindent as shown in Supplementary Fig.\,1. This corresponds to a parallel circuit. Note that the total trace (black curve) does not cross the 50\% line at the same fields as the separate traces (blue, green, and red lines). Not using such a model will wash out the transitions between adjacent superconducting phases.

In fitting the $\rho(B)$ isotherms of \iYRS\ (Fig.\,\ref{fig3}b), only for the
lowest temperature (0.6\,mK) the fit fully converged with all fit parameters
($k_i$, $B_{\rm{c2},i}$, $i = {\rm I}$, ${\rm II}$-${\rm left}$, ${\rm
II}$-${\rm right}$) open. Thus, to enable fits also at higher temperatures, we
used the $B_{\rm{c2,I}}$ value determined in this lowest-temperature fit and the
critical temperature in zero field to approximate the phase boundary with the
mean-field form $B_{\rm{c2}}(T)=B_{\rm{c2}}(T=0)\cdot [1-(T/T_{\rm c})^2]$, and
fixed $B_{\rm{c2,I}}$ in the higher-temperature fits to the corresponding
values. All other parameters were left open and the fits converged.

Also the low-temperature $\rho(B)$ isotherms of \YRS\ (Fig.\,\ref{fig3}a) show `double transition' signatures, but because they are less pronounced than in 
\iYRS\ we have refrained from doing a similar analysis there (the critical fields are simply determined by the 50\% resistivity drop criterion, as explained in the main text).
\vspace{0.5cm}

\noindent{\bf Supplementary Note~2: Estimates on Planckian dissipation}

\noindent In a material with Planckian dissipation, a linear-in-temperature electrical resistivity arises when the scattering rate $1/\tau$ reaches the Planckian limit, $k_{\rm B}T/\hbar$. To test whether this is the case in \YRS, we have determined the proportionality coefficient $\alpha$ in
\vspace{-0.5cm}

\begin{equation}
\frac{1}{\tau} = \alpha \frac{k_{\rm B}T}{\hbar} \label{planck}
\end{equation}

\noindent in two different ways. In a first approach, we use the simple Drude form
\vspace{-0.5cm}

\begin{equation}
\rho = \frac{m}{n e^2} \frac{1}{\tau} \label{Drude}
\end{equation}

\noindent for the inelastic part\vspace{-0.5cm}

\begin{equation}
\rho_{\rm in} = A' T \label{rho_dc}
\end{equation}

\noindent of the strange-metal electrical resistivity $\rho = \rho_0 + \rho_{\rm in}$, identify $1/\tau$ in Supplementary Eq.\,\ref{Drude} with Supplementary Eq.\,\ref{planck}, and obtain
\vspace{-0.5cm}

\begin{equation}
\alpha = \frac{n e^2}{m} \frac{\hbar}{k_{\rm B}}A' \label{alpha} \; .
\end{equation}

\noindent Here $n$ is the charge carrier concentration and $m$ the charge carriers' effective mass. Because \YRS\ is a multiband conductor \cite{Fri10.1}, estimating $n$ from Hall effect measurements \cite{Pas04.1,Fri10.2} in a simple one-band model may introduce a sizable error. Thus, instead, we determined $n$ from the Ginzburg-Landau coherence length $\xi_{\rm GL}$, the superconducting transition temperature $T_{\rm c}$, and the Sommerfeld coefficient $\gamma$, via the relation\vspace{-0.5cm}

\begin{equation}
n = \left( \frac{\xi_{\rm GL} T_{\rm c} \gamma}{7.95 \cdot 10^{-17}}\right ) ^{3/2} \quad \mbox{ in cgs units} ,
\end{equation}

\noindent as described in refs.\,\citenum{Rau82.1,Orl79.1}. $\xi_{\rm GL}$ and $T_{\rm c}$ are taken from Table~\ref{tab_sc}. $\gamma$ is calculated from the zero-field $A$ coefficient of the Fermi liquid form $\rho = \rho_0 + AT^2$ (from Table~\ref{tab_sample}) via the universal Kadowaki-Woods ratio $A/\gamma^2 = 10^{-5}\,\mu\Omega{\rm cm}\rm{(mol K)}^2/\rm{(mJ)}^2$. We expect this value, $\gamma_{\rm KW}^{\rm 0\,T}$ (see Table~\ref{tab_sample}) to be a more reliable estimate of the Sommerfeld coefficient than the directly measured specific heat data because the phase transition anomaly from the N\'eel transition \cite{Geg02.1,Kne06.1} is superimposed on the data and difficult to subtract. The corresponding averaged Fermi wavevector
\vspace{-0.5cm}

\begin{equation}
k_{\rm F} = (3 \pi^2 n)^{1/3} \label{kF}
\end{equation}

\noindent is given in Table~\ref{tab_sc} for both \YRS\ and \iYRS. This $k_{\rm
F}$ and $\gamma_{\rm KW}^{\rm 0\,T}$ are used to calculate the effective mass as
\vspace{-0.5cm}

\begin{equation}
m = \frac{3 \hbar^2}{k_{\rm B}^2} \frac{\gamma_{\rm KW}^{\rm 0\,T}}{k_{\rm F}}\; . \label{kF}
\end{equation}

\noindent The $\alpha$ values thus obtained from Supplementary Eq.\,\ref{alpha} are 0.011 and 0.0065 for \YRS\ and \iYRS, respectively (Table~\ref{tab_sc}), much smaller than $\alpha \approx 1$ expected for Planckian dissipation.

Our second approach makes use of recent optical conductivity measurements on \YRS\ \cite{Pro20.1}, which showed that the inverse of the real part of the inelastic optical conductivity (an optical resistivity) is linear in frequency $\nu$:\vspace{-0.5cm}

\begin{equation}
\frac{1}{{\rm Re}(\sigma_{\rm in})} = A'' \nu \; . \label{rho_ac}
\end{equation}

\noindent By equating Supplementary Eq.\,\ref{rho_dc} and Supplementary Eq.\,\ref{rho_ac} and using Supplementary Eq.\,\ref{planck} for $\nu = 1/\tau$ we can calculate $\alpha$ as\vspace{-0.5cm}

\begin{equation}
\alpha = \frac{\hbar}{k_{\rm B}} \frac{A'}{A''} \; , \label{alpha1}
\end{equation}

\noindent which yields 0.0062, in good agreement with the estimates from approach 1. 

\newpage

{\noindent\large\bf Supplementary Figures}
\vspace{0.8cm}

\begin{figure}[!h]
\begin{center}
\hspace{0.85cm}\includegraphics[height=0.36\textwidth]{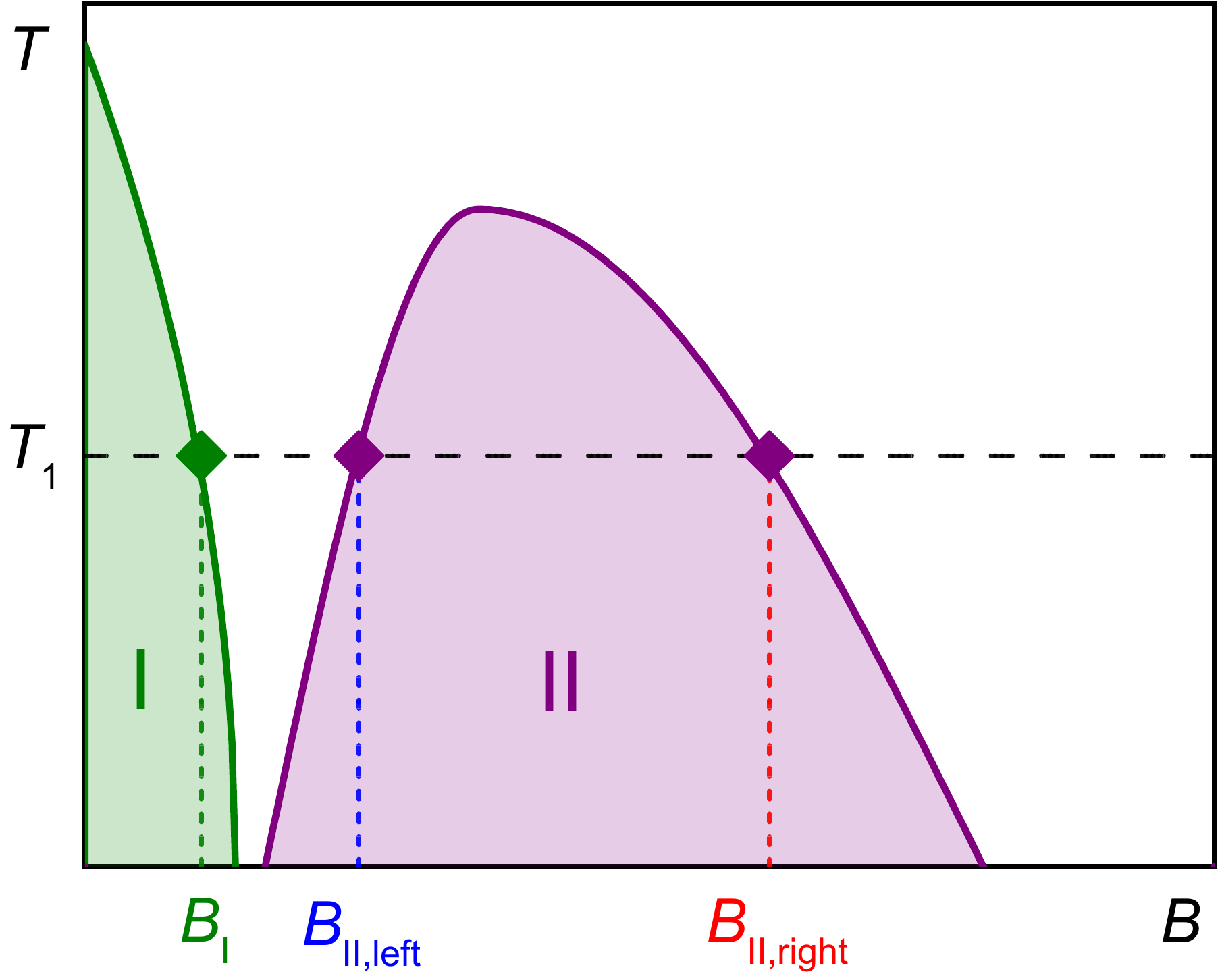}
\vspace{0.4cm}

\includegraphics[height=0.36\textwidth]{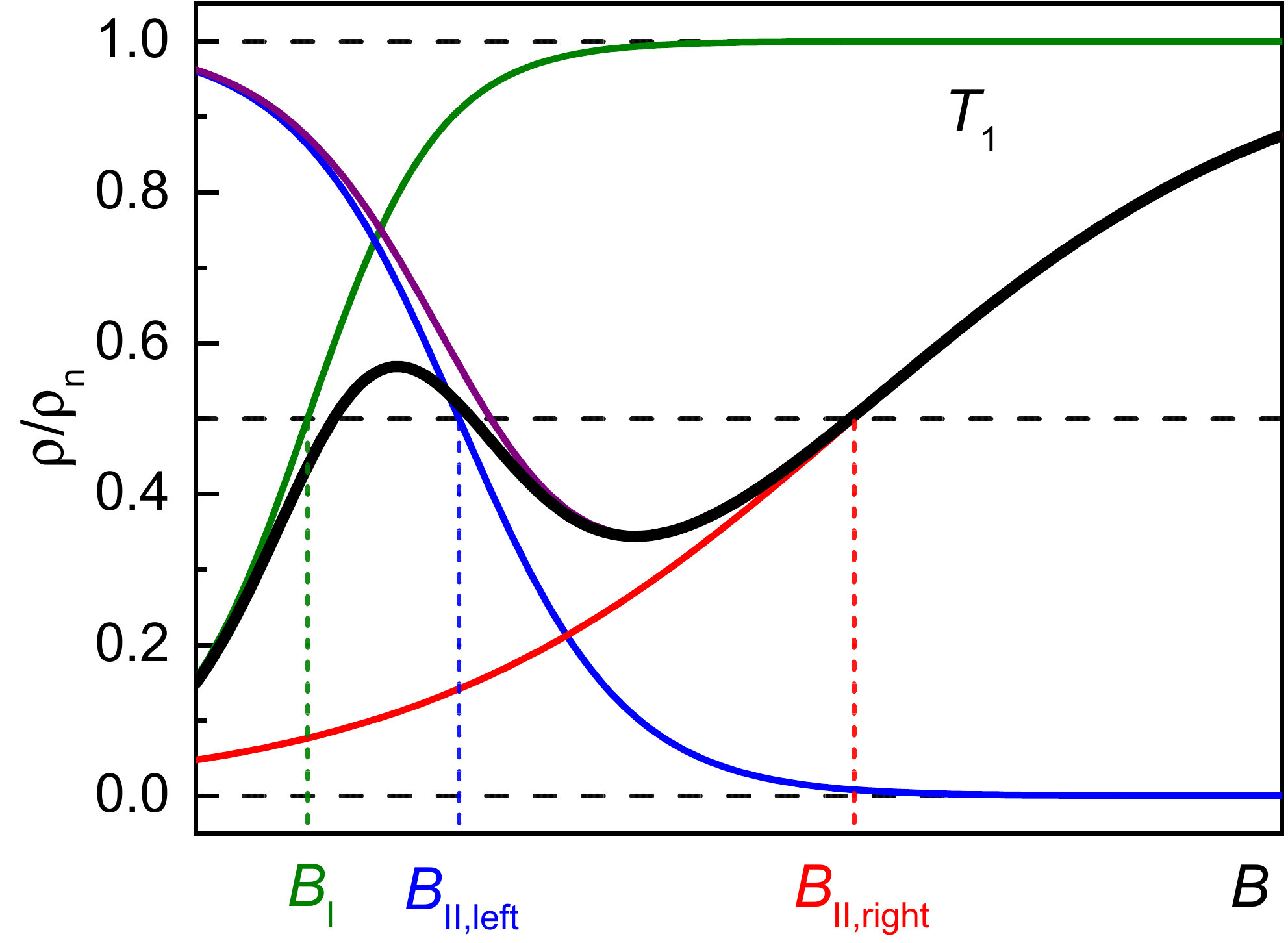}
\vspace{-13cm}

\hspace{-7.5cm}{\bf\textsf{a}}
\vspace{5.6cm}

\hspace{-7.5cm}{\bf\textsf{b}}
\vspace{5cm}
\end{center}

%\internallinenumbers%DELETE FOR PLAIN VERSION
%\caption{
\end{figure}
%\baselineskip24pt
%\nolinenumbers
\noindent {\bf Supplementary Fig.\,1: Determination of critical fields in a
conductor with two adjacent superconducting phases.} {\bf a} Example of two
superconducting phases, I and II, described by our simple `multi-transition
model' (see text). {\bf b} Electrical resistivity $\rho$, normalized to its
value in the normal state $\rho_{\rm n}$ just above the onset of
superconductivity, as function of the applied magnetic field $B$, at a fixed
temperature $T_1$. The total $\rho/\rho_{\rm n}(B)$ trace (black) is compose of
three logistic functions (green, blue, red), with transitions at $B_{\rm I}$,
$B_{\rm{II,left}}$, and $B_{\rm{II,right}}$, as described in the text.\\
%}
%\label{SMfig2}

%\bibliographystyleSM{naturemagallauthors}
%\bibliographySM{silke_all}
\newpage

\begin{figure}[!h]
\begin{center}
\includegraphics[width=0.75\textwidth]{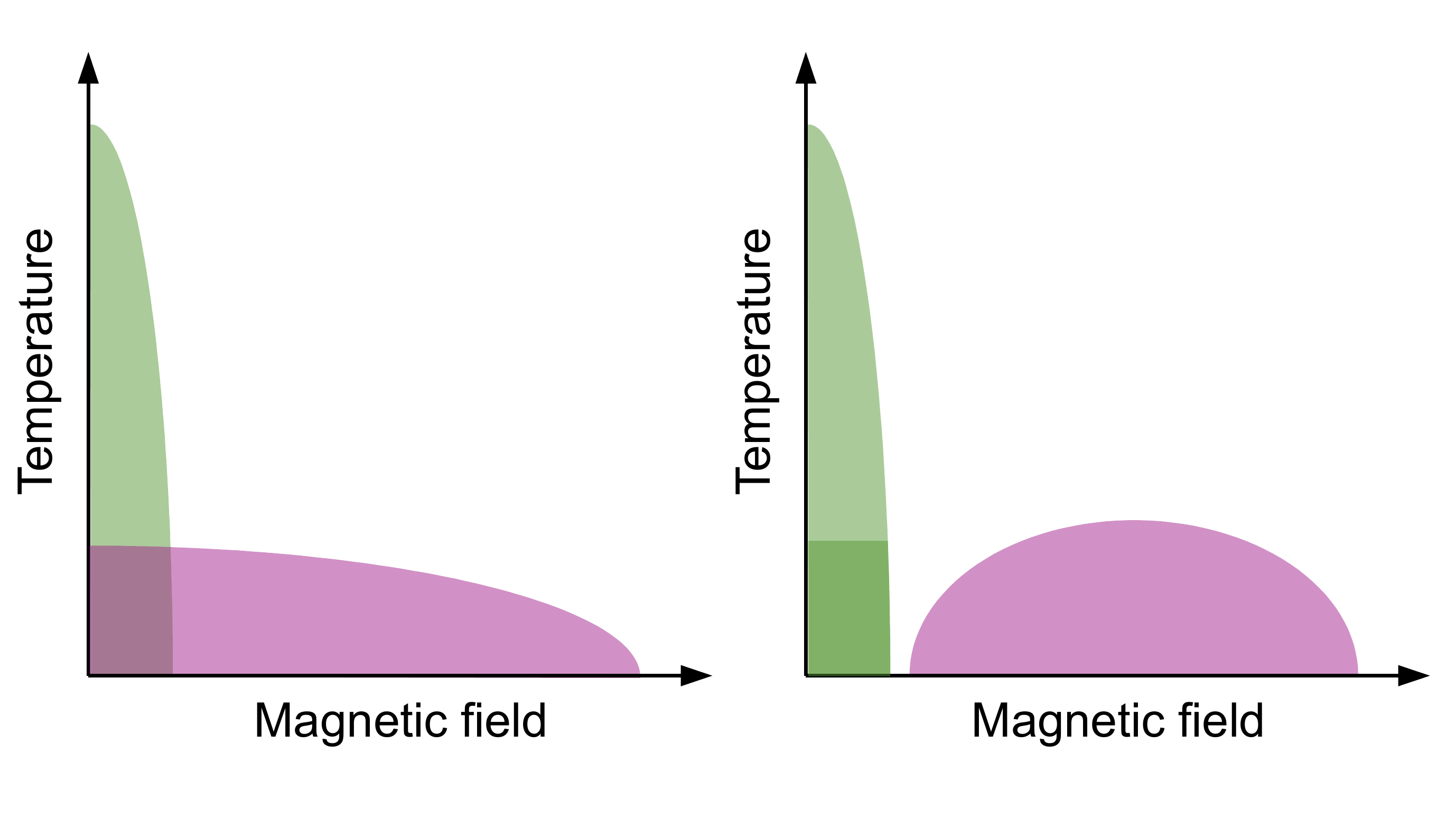}
\end{center}
\vspace{-7.5cm}

\hspace{-6cm}{\bf\textsf{a}}\hspace{5.9cm}{\bf\textsf{b}}
\vspace{6cm}
%\internallinenumbers%DELETE FOR PLAIN VERSION
%\caption{
\end{figure}
%\baselineskip24pt
%\nolinenumbers
\noindent {\bf Supplementary Fig.\,2: Cartoons of possible superconducting phases in \YRS.} {\bf a} Two intersecting phases. {\bf b} Two separated phases, with an internal phase boundary in the low-field phase. Both are consistent with the combined results from the present work and ref.\,\citenum{Sch16.1}.\\

\newpage

\begin{figure}[!h]
\begin{center}
\includegraphics[height=0.25\textwidth]{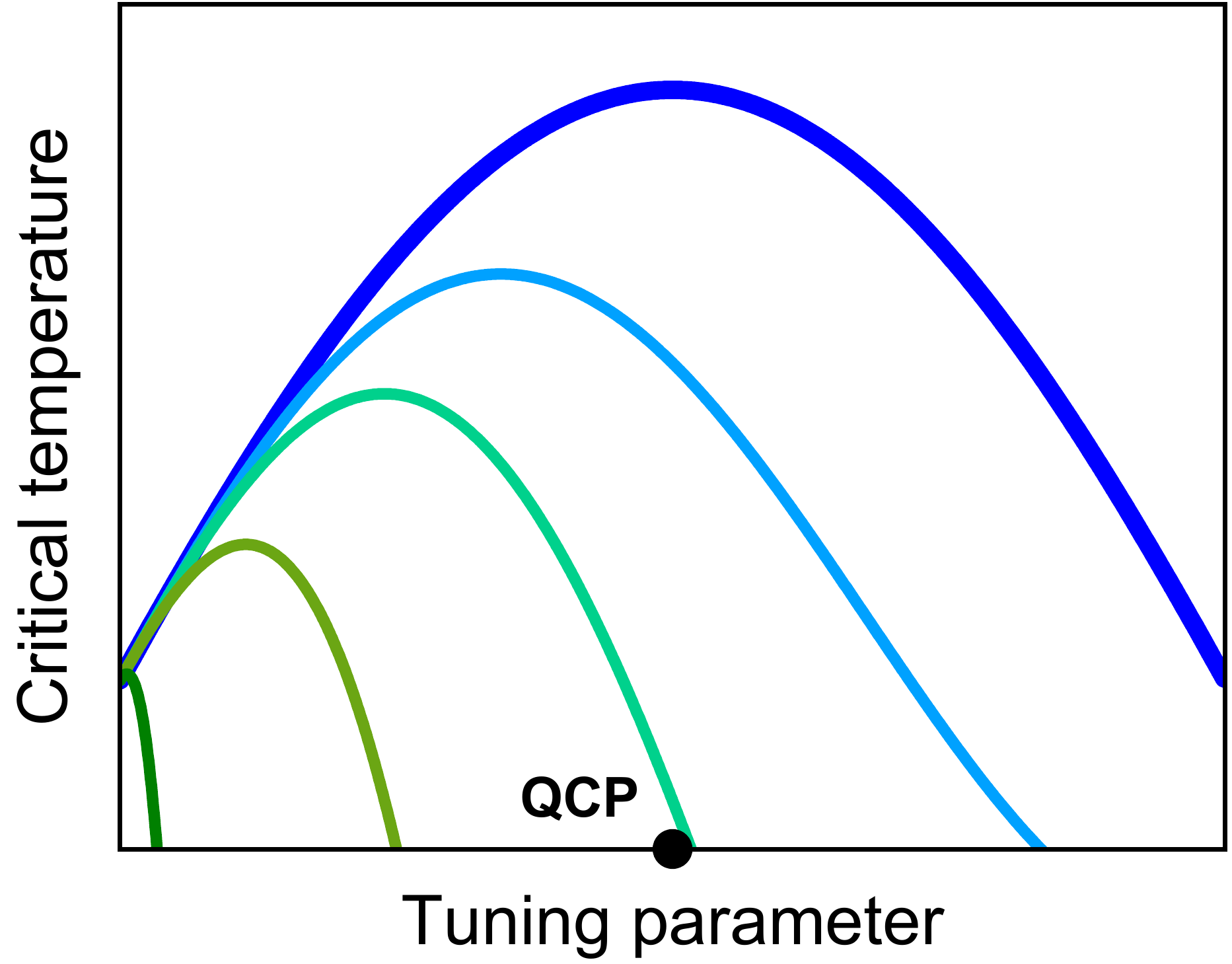}\hspace{0.1cm}
\includegraphics[height=0.25\textwidth]{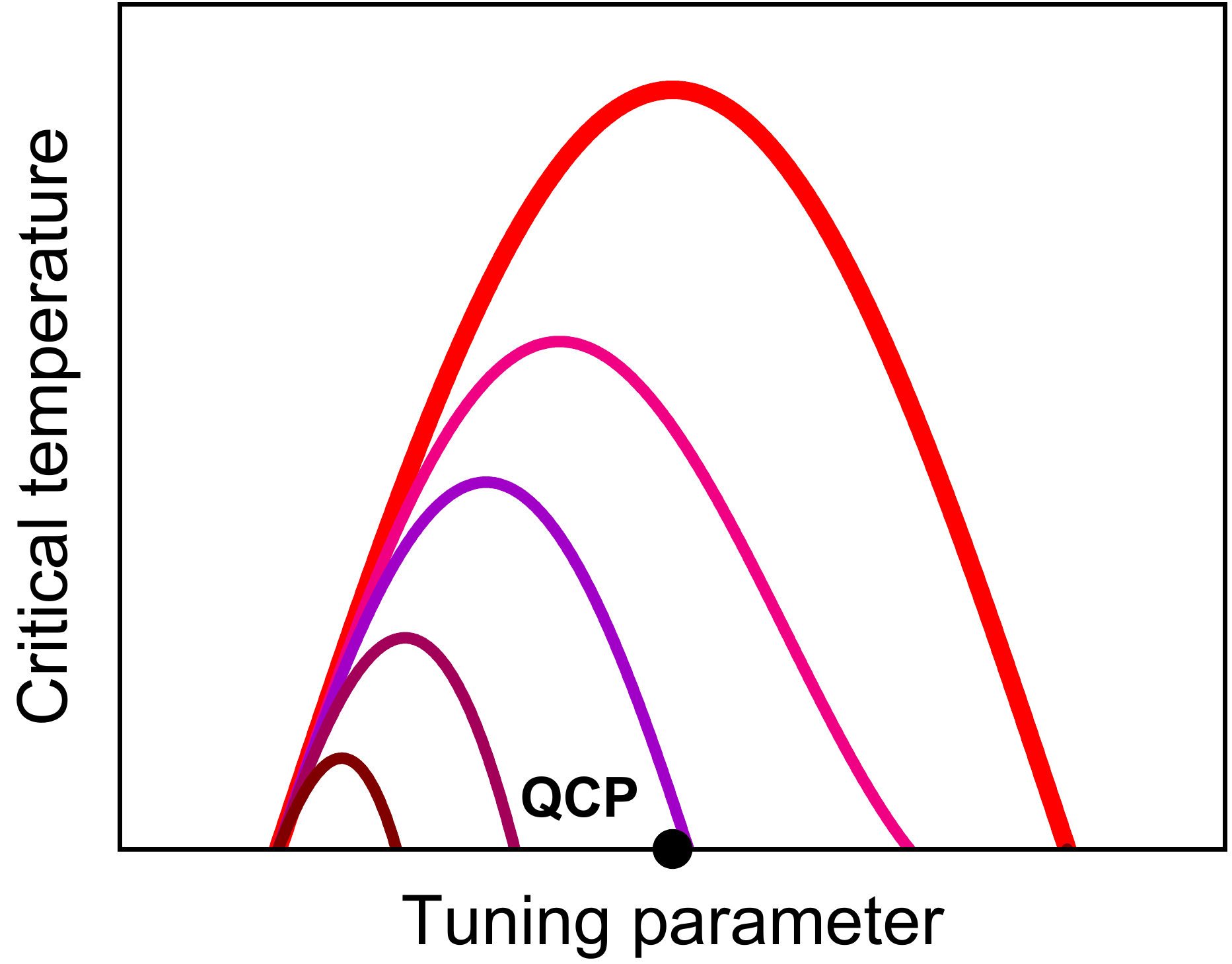}\hspace{0.1cm}
\includegraphics[height=0.25\textwidth]{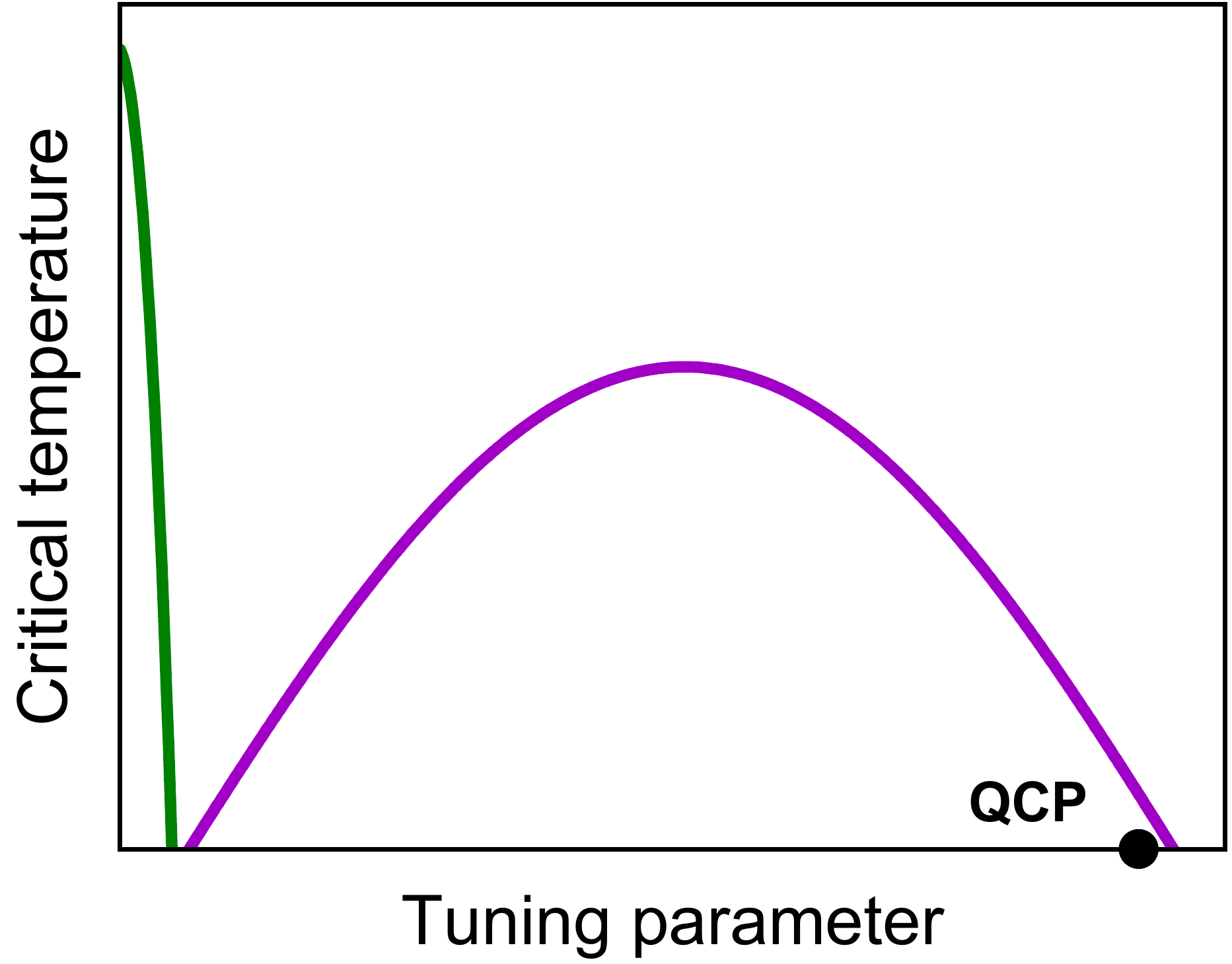}
\end{center}
\vspace{-5.2cm}

\hspace{-5cm}{\bf\textsf{a}}\hspace{5.2cm}{\bf\textsf{b}}\hspace{5.2cm}{\bf\textsf{c}}
\vspace{4cm}
%\internallinenumbers%DELETE FOR PLAIN VERSION
%\caption{
\end{figure}
%\baselineskip24pt
%\nolinenumbers
\noindent {\bf Supplementary Fig.\,3: Cartoons of magnetic-field effect on superconducting domes around a quantum critical point (QCP).} {\bf a} The fat blue curve is a hypothetical $T_{\rm c}(B)$ curve where any (hostile) effect of the magnetic field on superconductivity (from Pauli or orbital limiting) is imagined to be absent. For the successive curves (from blue to green), an increasingly strong field effect is considered (using a simple mean-field-type suppression of $T_{\rm c}$ by the magnetic field $B$, which increases from zero along the tuning parameter axis). {\bf b} Same as {\bf a}, but for a superconducting phase with a dome that does not extend to the zero of the tuning parameter axis (fat red curve). Again, an increasingly strong field effect is added for the successive curves (from red to brown). {\bf c} Lowest curve from panel {\bf a} (green) and middle curve of panel {\bf b} (purple), scaled in absolute values. A material with two QCP-derived superconducting phases, one with stronger pairing but larger field sensitivity (as in panel {\bf a}, e.g., a spin-singlet superconductor) and one with weaker pairing but also weaker field sensitivity (as in panel {\bf b}, e.g., a spin-triplet superconductor), could display such a magnetic-field-tuned phase diagram.

\end{document}